# Integration of a New Knobbox in the PSI Control Systems ACS and EPICS

A.C.Mezger, D.Anicic, T.Blumer, I. Jirousek
Paul Scherrer Institute, Switzerland

Abstract

Operation of the PSI cyclotrons has always relied heavily on manual control. Tuning of our high intensity beams for minimum losses is an art still depending on personal skill. An ergonomically good design of the knobbox is therefore essential for good operation. The age of the previous design, and the request, to implement the ACS knobbox in the SLS control system based on EPICS, was an opportunity for a redesign. The new knobbox is an autonomous node in the system. It is based on industrial pc104 hardware with LINUX as the operating system. CDEV and ACS-communication provide the interface to the respective control systems.

## 1 INTRODUCTION

For the operation of the PSI accelerators the Man-Machine Interface (MMI) to the power supplies, motors, etc consists of a panel presenting knobs, buttons and displays, the so-called knobbox (Fig. 1). The functionality of this interface has been optimized for the handling of our accelerators over the last 20 years. It has the ability to make small increments to a device through buttons, as well as continuous variation with knobs. Many other useful functions have been implemented for keeping the old value when a device is hooked, setting back this old value as well as the possibility to set a new value, switching the power supplies off and a HP-like calculator.

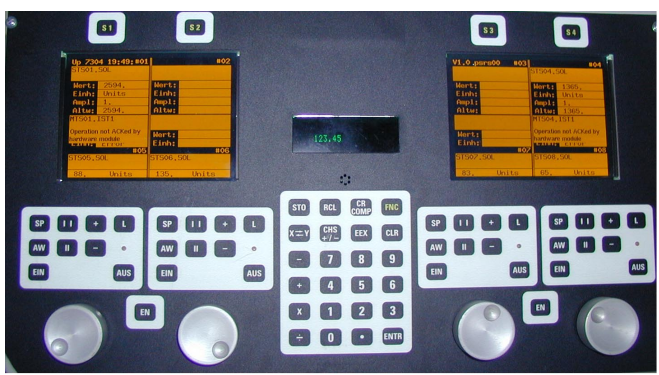

**Figure 1:** Knobbox MMI

## 2 HARDWARE EVOLUTION

The first generation of this hardware was implemented through CAMAC modules directly driven by a PDP control computer. The next generation saw the hardware controlled by a CAMAC auxiliary crate controller (ACC, CES 2180) and communicating with the control computer through the parallel CAMAC branch [1].

In the early nineties our control system evolved to a distributed system with front-end and back-end computers, all connected through an Ethernet network (Fig. 2).

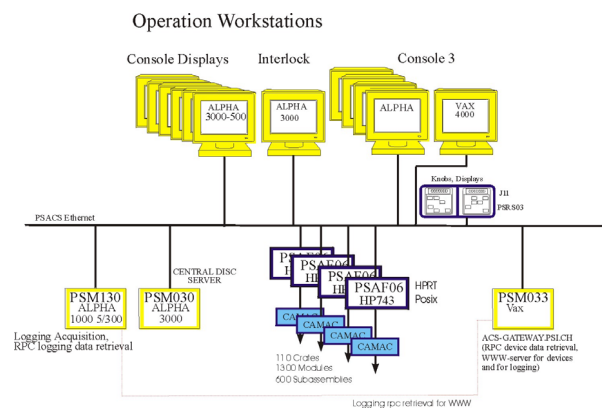

**Figure 2:** Control System layout

In this architecture, the front-end computers perform the data acquisition on the field bus (VME and CAMAC) and pass the results to the back-end (demanding) computer [2]. To integrate the knobbox into this architecture, a direct connection to the ACC, via an Ethernet interface, has been implemented. This made the crate at this moment standalone and its connection to the parallel CAMAC branch was removed.

## 3 NEW DEVELOPMENT

The knobbox is a necessary tool for the operation of our accelerators and cannot be replaced easily in all its functionality by modern software. Our Operators are accustomed to it and it is also used in the SLS (Swiss

Light Source) control system. We realized that the hardware associated to the knobbox is getting older, is expensive and furthermore that CAMAC is not always the preferred solution. We looked therefore for a new hardware solution where we could keep the same MMI.

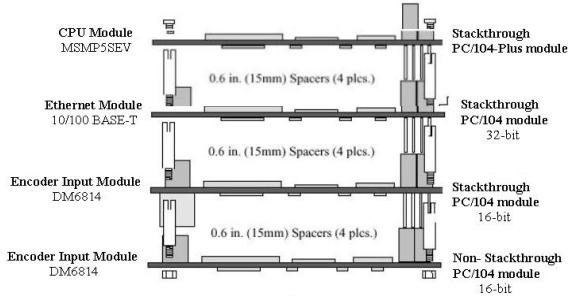

**Figure 3:** Typical module stack for the knobbox

The solution we have adopted uses an industrial embedded PC/104 system (smartCore-P5, a 266 MHz Pentium PC) with the necessary IO-cards: Up-Down Counters for the knobs, digital IO and Ethernet 10/100 (Fig. 3). These PC/104 cards are much smaller than ISA/PCI-bus cards found in PC's and stack together which eliminates the need for a motherboard, back plane, and/or card cage. Power requirements and signal drive are reduced to meet the needs of an embedded system. Because PC/104 is essentially a PC with a different form factor, any popular operating system can be used for this system. We have chosen LINUX.

The 10-year-old displays of the original knobbox have been replaced with quarter-VGA electro luminescent displays, giving the possibility to drive the two displays through the TFT connector of the PC/104. The small character display situated in the middle of our knobbox is driven through the parallel port of the PC/104. To simplify the handling of the keyboard and lights we used a field-programmable gate array from XILINX interfaced to the IO-cards. This layout is shown in Fig.4.

## 4 SOFTWARE IMPLEMENTATION

Because of the new architecture of the knobbox, the software has had to be completely rewritten (the old software was written using FORTRAN and was based on CAMAC hardware). The new software is completely written in C and does its IO through the IO-ports of the PC/104 system. The software uses X-WINDOWS for displaying on the "quarter-VGA" screens. To support these screens the PC/104 BIOS had to be modified; this was done by the Company that supplied us the PC/104 system. The data-acquisition is implemented in the standard way as defined by our control system, however CDEV (CDEV is a C++ framework for developing portable control applications developed at Thomas Jefferson Laboratory) has also been implemented. This extension to the knobbox allows it to be used directly with any system using CDEV. This requirement is important in the case of SLS or even if several control systems have to be addressed by the knobbox.

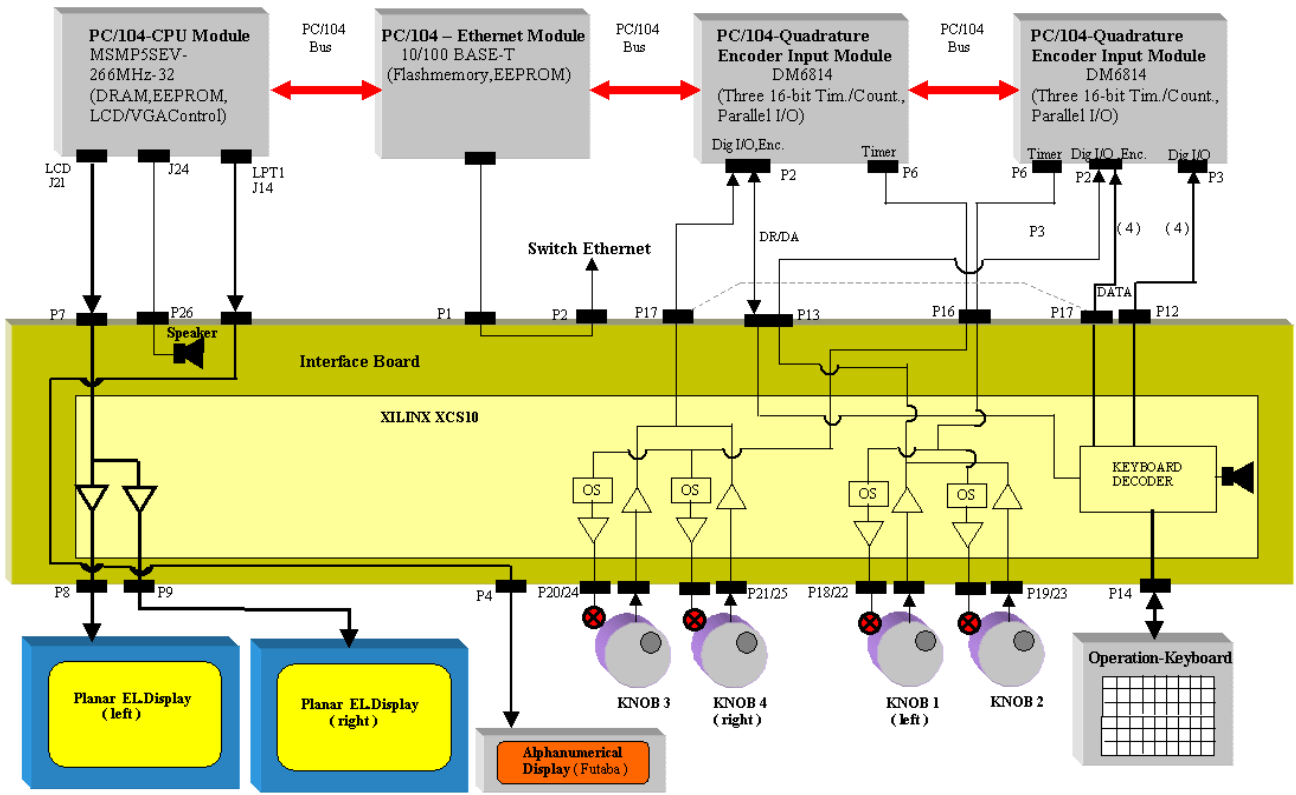

**Figure 4:** Knobbox block diagram system.

For the development of our knobbox, a small laptop disk, having the Linux system and the necessary software, was connected to the PC/104. Recently we changed to a diskless system booted from a server, so that several knobboxes can be installed with only a small effort.

Two knobboxes are actually in the test phase and will soon be installed into our Control room, so that we can test the knobbox under real conditions.

## 5 ACKNOWLEDGMENTS

The authors would like to thank the following people for their contributions to the hardware of the upgraded knobbox: D.Buerki, W.Hugentobler and G.Janser.